\documentclass[12pt]{article}
\usepackage{amsmath}
\usepackage{amsfonts}
\usepackage{amssymb}
\usepackage{graphicx}
\newcommand*{\Tr}{\mathrm{Tr}}
\begin{document}

\begin{titlepage}
\ \\
\begin{center}
\LARGE
{\bf
Channel Estimation Theory of Low-Noise Multiple Parameters:\\
Attainablity Problem of the Cram{\'e}r-Rao Bounds
}
\end{center}
\ \\
\begin{center}
\large{
M. Hotta${}^{\ast}$
\footnote[1]{hotta@tuhep.phys.tohoku.ac.jp}
and T. Karasawa${}^{\dag}$ 
\footnote[2]{jidai@nii.ac.jp}
}\\
{\it
${}^\ast$
Department of Physics, Faculty of Science, Tohoku University,\\
Sendai, 980-8578, Japan\\
${}^{\dag}$
National Institute of Informatics,\\
2-1-2 Hitotsubashi, Chiyoda-ku, Tokyo, 101-8430, Japan }
\end{center}

\begin{abstract}
For decoherence processes induced by weak interactions with the environment, 
a general quantum channel with one noise parameter has been formulated. This channel is called low-noise channel and very useful for investigating the parameter estimation in the leading order. In this paper, we formulate the low-noise channel with multiple unknown parameters in order to address the simultaneous achievability of the Cram{\'e}r-Rao bound for the parameters estimation. In general, the simultaneous achievement of the Cram{\'e}r-Rao bound for multi-parameter estimations suffers from non-commutativity of optimal measurements for respective parameters.  However, with certain exceptions,  we show that the Cram{\'e}r-Rao bound for output states of dissipative low-noise channels can be always 	attained in the first order of the parameters as long as  $D\leq N-1$, where $D$ and $N$ denote the number of the parameters and the dimension of the system, respectively. This condition is replaced by $D\leq N^{2}-1$ if it is allowed to set the entanglement with ancilla systems in its input state and to perform the non-local measurement on the composite system.
\end{abstract}
\end{titlepage}

\section{Introduction}
\

Decoherence processes driven by weak interactions with the environment can be described in a unified way by the low-noise channel formulation proposed in Ref.~\cite{hko}, which has a wide range of applications for the estimation of small noise parameters on quantum channels. An important example of its application is the noise estimation in quantum computing~\cite{NC} in which small noises induced by the environment can lead to a serious obstacle to its implementation. One of the best way to eradicate the noise effect is to apply an appropriate error correcting code to the system. In this scheme, one first needs to know properties of the noises to design the code. However, because the noises are generated by very complicated many-body interactions, it is difficult to calculate the properties theoretically from first principles. Therefore, it is effective for finding out the properties to measure the noise experimentally and to estimate them from the experimental data. With this method, the estimation of such very weak noises  may suffer from large ambiguity of experimental errors. Thus the estimation theory of the low-noise channel will play a significant role in order to improve the performance of the estimation. Another example to which the estimation theory of the low-noise channel can be applied
is found in rare processes of the elementary particle physics~\cite{K}. 
Many theories of new particles has predicted the existence of very weak interactions which generate rare decays and rare reactions of particles. However, in real experiments, the number of signals indicating new physics is generally small, compared with that of
standard-model signals. Thus, siding with the standard model, the evidential
signals can be regarded as low-noise background datum. The estimation theory
of the low-noise channel is expected to improve accuracy of estimation of
such tiny reaction rates and to assist in the finding of new physics.

An estimation theory for the low-noise channel has been formulated for one noise parameter, termed a low-noise parameter, which is assumed to be unknown and very small~\cite{hko}.  The increase of the Fisher information due to the prior entanglement with non-local output measurement, which is called the ancilla-assisted enhancement, has been analyzed for the ancilla-extended version of the low-noise channel~\cite{hko}. The low-noise channel has been also extended to the channel on the many-body system~\cite{hko2}, and this work shows that the maximum of the Fisher information can be attained by a factorized input state.

In this paper, we address the estimation problem of multiple low-noise parameters $\epsilon =(\epsilon ^{1},\cdots ,\epsilon ^{D})$, where $D$ is the number of the parameters.  We assume that all of $\epsilon ^{\mu }$ are non-negative and take very small values of the same order. In later analysis, $O(\epsilon )$ denotes the order of $\epsilon ^{\mu }$, and the orders of products of the parameters, $\prod_{\mu =1}^{D}(\epsilon ^{\mu })^{n_{\mu}}$, are denoted by $O\left( \epsilon ^{n_{1}+\cdots +n_{D}}\right) $. Due to the multi-parameter degrees of freedom, a large variety of physical phenomena can be analyzed by use of this low-noise channel.

The low-noise channel are dependent on the unknown parameters $\epsilon$ which are estimated by measurements. Consider an input state $\rho _{in}$, which is independent of $\epsilon$, into the low-noise channel $\Gamma_{\epsilon}.$ The output state $\rho =\rho _{out}(\epsilon)$ is obtained by 
\begin{equation}
\rho _{out}(\epsilon )=\Gamma _{\epsilon }[\rho _{in}].  \label{q1}
\end{equation}
If the purity of the output states is decreased, then the low-noise channel is called dissipative. In what follows, we focus on the dissipative low-noise channel.

To estimate the low-noise parameters, a POVM measurement with $D$ measurement values are performed on the output state $\rho _{out}(\epsilon )$.  The measurement values are regarded as estimation values for $\epsilon ^{\mu }$ and this POVM is called an estimator. When the expectation value of the estimation value is equal to the true value of the parameter $\epsilon ^{\mu}$ for any $\mu$, the estimator is called locally unbiased.   In the present paper the locally unbiased estimator is defined in the vicinity of $\epsilon =0$. Note that by virtue of the Na\^{\i}mark theorem~\cite{hels,holevo}, the POVM measurement is equivalent to a measurement of commutative observables (projection-valued measures) in a composite system comprising the original system and an ancilla system.

One of the fundamental quantities in the low-noise estimation theory is the
Fisher information matrix of an output state of the low-noise
channel $\Gamma _{\epsilon }$. It is well known as the Cram\'{e}r-Rao inequality that the inverse of the Fisher information matrix is a lower bound of the mean-square-error matrix of the locally unbiased estimators \cite{hels,holevo}. In this paper, we focus our
attention on the achievement problem of the lower bound for the dissipative
low-noise channel. In the case with one low-noise parameter, the bound can
be attained by using certain estimators. In contrast, the bound cannot be always achieved if $D>1$. Even if a certain unbiased estimator is optimal for the parameter estimation of $\epsilon ^{1}$, the estimator is not always optimal for other parameters $\epsilon^{2},\cdots ,\epsilon ^{D}$, because of the non-commutativity of optimal estimators.  It is an important open problem to reveal the necessary and sufficient condition for the attainability of the bound for the multi-parameter
estimation. The solution for general channels has not been found to date.
Recent progress has been reported on the multi-parameter estimation for
special families of quantum channels, in particular, SU(N) channels \cite{F02,if,Kahn} and a generalized Pauli channel \cite{FI}.

Our main results are related to the attainability problem for the estimation in the dissipative low-noise channel. To explain the results, let us diagonalize the output state $\rho(\epsilon)$ as
\begin{equation*}
\rho (\epsilon )=p_{0}(\epsilon )|0(\epsilon )\rangle \langle 0(\epsilon
)|+\sum_{n=1}^{N-1}p_{n}(\epsilon )|n(\epsilon )\rangle \langle n(\epsilon
)|,
\end{equation*}
where the first eigenvalue $p_{0}(\epsilon )$ is $O(\epsilon ^{0})$, the
other eigenvalues $p_{n\neq 0}(\epsilon )$ are $O(\epsilon )$, and $N$ denotes the dimension of the Hilbert space of the quantum system (the reason why this expression is available is shown in Sec.~\ref{channel-estimation}). The Fisher information of the output state for the channel has a leading term of order of $O(\epsilon ^{-1})$. 
We first prove in the case with $D\leq N-1$ that all output states of the dissipative low-noise channel have locally unbiased estimators by which the lower bound is attained in $O(\epsilon)$ assuming that 
\begin{equation}
\det_{\mu \nu }\left[ \sum_{n=0}^{N-1}\partial _{\mu }\sqrt{p_{n}(\epsilon )}\partial _{\nu }\sqrt{p_{n}(\epsilon )}\right] \neq 0  \label{100}
\end{equation}
in the vicinity of $\epsilon =0$, where  $\partial _{\mu }$ stand for $\partial/\partial \epsilon^{\mu}$. 
In order to explain the meaning of Eq.~(\ref{100}), let us consider a map 
$( \sqrt{p_{0}(\epsilon )},\cdots ,\sqrt{p_{N-1}(\epsilon )}) $ which generates a manifold by running the $D$ parameters of $\epsilon $. Clearly, if $D>N-1$, Eq.~(\ref{100}) does not
hold because the target space defined by $$\left\{ \left( \sqrt{p_{0}},\cdots
,\sqrt{p_{N-1}}\right) \Big| p_{i}\geqq 0,\sum_{i=0}^{N-1}p_{i}=1\right\} $$ has $N-1$ dimensions and the parameterization of the manifold by $\epsilon $ becomes degenerate. On the other hand, when $D\leq N-1$, it is noticed that
imposition of Eq.~(\ref{100}) is not so difficult. If the dimension of the
manifold is equal to $D$, the relation in Eq.~(\ref{100}) automatically holds.
Therefore, the condition in Eq.~(\ref{100}) is not anomalous and can be
satisfied for many standard low-noise channels.

Next, we point out that, although the lower bound is not always achieved
even in $O(\epsilon ^{0})$ when $D>N-1$, the entanglement with ancilla systems can make the lower bound attainable again in $O(\epsilon )$ when $N^{2}-1\geq D$. As is expected from the fact that the dissipative low-noise channel comprise a rather general class of quantum channels, these results are very surprising because they reveals that an enormous number of channels can overcome possible non-commutativity among the optimal estimators to attain the Cram\'{e}r-Rao lower bound.

\section{Brief Review of State-Parameter Estimation \label{review}} 
\ 

We start with a brief review of a parameter estimation theory of quantum states,  since the channel estimation theory is based on the state estimation theory.
Let us consider a state $\rho $ dependent on $D$ unknown parameters $\theta ^{\mu},$ 
\begin{equation}
\rho =\rho (\theta )=\rho (\theta ^{1},\cdots ,\theta ^{D}).
\end{equation}
 The parameters $\theta^{\mu }$ are estimated by measuring the state $\rho $ with a positive operator-valued measure (POVM) $\left\{ \Pi _{x}|x=(x^{1},\cdots ,x^{D})\right\}$ which satisfies
\begin{equation}
\Pi _{x}\geq 0, \hspace{2mm} \int \Pi _{x}d^{D}x=\mathbf{1,}
\end{equation}
by definition. The POVM $\Pi _{x}$ is called the estimator for the parameter
estimation. When the condition 
\begin{equation}
\int x^{\mu }\Tr[\Pi _{x}\rho (\theta )]d^{D}x=\theta ^{\mu }
\end{equation}
is satisfied, the estimator is called unbiased. For the unbiased estimator, the mean-square-error matrix $V^{\mu \nu }$ is
defined by
\begin{equation}
V^{\mu \nu }[\Pi _{x}]:=\int (x^{\mu }-\theta ^{\mu })(x^{\nu }-\theta ^{\nu
})\Tr[\Pi _{x}\rho (\theta )]d^{D}x. 
\end{equation}
As is well known, there exists a universal lower bound of the error matrix $V^{\mu \nu}$, called the Cram\'{e}r-Rao bound~\cite{hels,holevo}, as 
\begin{equation}
V\geq J^{-1}, \label{Cramer-Rao}
\end{equation}
where $J$ stands for the Fisher information matrix for $\rho$ defined by
\begin{equation}
J_{\mu \nu }=\frac{1}{2}\Tr[\rho (L_{\mu }L_{\nu }+L_{\nu }L_{\mu })],
\end{equation}
and $L_{\mu}$'s are the symmetric logarithmic derivative (SLD) defined by
\begin{equation}
L_{\mu }^{\dagger }=L_{\mu }, \hspace{1mm} \frac{\partial }{\partial \theta ^{\mu }}\rho = \frac{1}{2}(L_{\mu }\rho +\rho L_{\mu }).
\end{equation}
Eq.~(\ref{Cramer-Rao}) is called the Cram\'{e}r-Rao inequality and 
 implies that for an arbitrary $D$-dimensional vector $\overrightarrow{u}
$, the following relation is always satisfied, 
\begin{equation}
\overrightarrow{u}V\overrightarrow{u}\geq \overrightarrow{u}J^{-1}\overrightarrow{u}.  \label{1}
\end{equation}

Note that the equality is not always attainable. This aspect is related to
the non-commutativity of quantum measurements. When $D=1$, measurement of
the SLD operator $L_{1}$ is optimal for $\theta ^{1}$ estimation, that is,
the equality $V=J^{-1}$ is satisfied. When $D=2$, each measurement of the
SLD operator $L_{\mu }$ is optimal for $\theta ^{\mu }$ estimation
independently. However, the observables $L_{1}$ and $L_{2}$ do not
generally commute to each other, 
\begin{equation*}
\lbrack L_{1},L_{2}]\neq 0.
\end{equation*}
This non-commutativity of SLD's makes achievement conditions of the bound
nontrivial. For the pure-state case ($\rho (\theta )=|\phi (\theta )\rangle
\langle \phi (\theta )|$), it has been proven in Ref.~\cite{m1} that the bound can be always attained. However, generally, the necessary and sufficient condition for the bound achievement has not been known.

The state estimation theory explained thus far can be applied to the channel
estimation done as follows.  Let us consider a $\theta ^{\mu }$-parameterized channel $\Gamma_{\theta }$. The channel generates an output state $\rho (\theta )=\Gamma_{\theta }[\rho _{in}]$ from a input state $\rho _{in}$ independent of $\theta ^{\mu }$.  The parameter $\theta ^{\mu }$ on the channel is estimated by use of the state estimation theory for the output state $\rho (\theta )$.

The Fisher information matrix $J[\rho _{in}]$ for the output state changes as the input state $\rho_{in}$ changes. Hence, it is important to find input states which optimize $J$. For this problem, there is a useful theorem \cite{f1} related to choice of the input state. There exists a pure input state $|\phi \rangle \langle \phi |$ for which the Fisher information matrix $J$ is the maximum over all input states $\rho _{in}$, that is, for an arbitrary vector $\overrightarrow{u}$,the following relation holds.
\begin{equation}
\overrightarrow{u}J[|\phi \rangle \langle \phi |]\overrightarrow{u}\geq 
\overrightarrow{u}J[\rho _{in}]\overrightarrow{u}.
\end{equation}
As was seen in Eq.(\ref{Cramer-Rao}), the mean-square-error matrix $V$ is lower bounded by the inverse of $J$, therefore, without loss of generality, we may concentrate on pure input states in later discussions.

As mentioned above, the Cram\'{e}r-Rao bound for the output states cannot be always
attained and the necessary and sufficient condition for its achievement in
channel estimation is a crucial open problem. For unitary channels, the achievability problems of the bound can be completely solved by use of the result in Ref.~\cite{m1}, since output states of the channels for pure input states remain pure, and thus the estimation problem of the channels is reduced to that of the output pure states. To date, the solution of the
lower-bound achievement has not been found for general channels.

\section{\protect\bigskip Multi-Parameter Low-Noise Channel \label{low-noise-channel}}
\ 

In this section, we define the low-noise channel $\Gamma _{\epsilon }$ with unknown multi-parameter $\epsilon$.
This channel is a natural extension of the low-noise channel with one parameter first introduced in Ref. \cite{hko}. The definition is given by a Kraus representation as
\begin{equation}
\rho (\epsilon )=\Gamma _{\epsilon }[\rho ]=\sum_{\alpha }B_{\alpha
}(\epsilon )\rho B_{\alpha }^{\dagger }(\epsilon )+\sum_{\mu =1}^{D}\epsilon
^{\mu }\sum_{a=1}^{K_{\mu }}C_{\mu a}(\epsilon )\rho C_{\mu a}^{\dagger
}(\epsilon ),  \label{lnc}
\end{equation}
where the Kraus operators satisfy the following four conditions.

(i) The channel is a TPCP map: 
\begin{equation}
\sum_{\alpha }B_{\alpha }^{\dagger }(\epsilon )B_{\alpha }(\epsilon
)+\sum_{\mu =1}^{D}\epsilon ^{\mu }\sum_{a=1}^{K_{\mu }}C_{\mu a}^{\dagger
}(\epsilon )C_{\mu a}(\epsilon )=\mathbf{1}_{S}.  \label{cp}
\end{equation}

(ii) $B_{\alpha}(\epsilon)$ is analytic at $\epsilon=0$, giving the power
series expansion  
\begin{equation}
B_{\alpha}(\epsilon)=\kappa_{\alpha}\mathbf{1}_{S}-\sum_{(n_{1,\cdots,}n_{D})\neq(0,\cdots,0)}N_{\alpha}^{(n_{1}\cdots
n_{D})}(\epsilon^{1})^{n_{1}}\cdots(\epsilon^{D})^{n_{D}},
\end{equation}
in the neighborhood of $\epsilon=0$, where $\kappa_{\alpha}$ and $N_{\alpha}^{(n_{1}\cdots n_{D})}$ are coefficients and operators,
respectively, independent of $\epsilon$.

(iii) $\kappa _{\alpha }$ satisfies 
\begin{equation*}
\sum_{\alpha} |\kappa _{\alpha }|^{2}=1.
\end{equation*}

(iv) $C_{\mu a}(\epsilon )$ is analytic at $\epsilon =0$, giving the power
series expansion 
\begin{equation}
C_{\mu a}(\epsilon )=M_{\mu a}+\sum_{(n_{1,\cdots ,}n_{D})\neq (0,\cdots
,0)}M_{\mu a}^{(n_{1}\cdots n_{D})}(\epsilon ^{1})^{n_{1}}\cdots (\epsilon
^{D})^{n_{D}},  \label{2}
\end{equation}
where $M_{\mu a}$ and $M_{\mu a}^{(n_{1}\cdots n_{D})}$ are operators independent of $\epsilon $.

The condition (i) is a natural characteristic of physical channels, because all physical channels are TPCP maps and any TPCP map has Kraus representations. The conditions (ii) and (iv) simply imply that the channel shows nonsingular behavior near $\epsilon =0$. Therefore, taking proper limits of 
weak coupling with the environment, physical processes induced by the
environment can always be described by this low-noise channel. From condition (iii), the channel automatically reduces to the identical
channel in the limit of vanishing the parameters:
\begin{equation}
\lim_{\epsilon \rightarrow +0}\Gamma _{\epsilon }=id_{S}.  \label{id},
\end{equation}
where $id_{S}$ stands for the identity channel on the system $S$. This shows that the parameters $\epsilon^{\mu}$ can represent noise parameters caused by the environment.

 The second term of the right-hand-side of Eq.~(\ref{lnc}) describes a dissipative effect by the environment in the lowest order of $\epsilon$ and plays a crucial role in the later discussion. Without loss of generality, we may assume that each of $C_{\mu a}$ is not proportional to others. If not, the second term can be rewritten in the
form of Eq.~(\ref{lnc}) with a smaller number of independent $C_{\mu
a}(\epsilon )$ by using coordinate transformation of $\epsilon ^{\mu }$. For the sake of convenience,  
let us denote $K$ the total number of $C_{\mu a}(\epsilon )$ as
\begin{equation*}
K=\sum_{\mu =1}^{D}K_{\mu }.
\end{equation*}
In addition, we redefine
the Kraus operators as
\begin{equation}
N_{1a}=N_{\alpha}^{(100\cdots0)},N_{2\alpha}=N_{\alpha}^{(010\cdots0)}, \cdots,N_{D\alpha}=N_{\alpha}^{(00\cdots01)}.
\end{equation}
Then $B_{\alpha}(\epsilon)$ is rewritten as
\begin{equation}
B_{\alpha }(\epsilon )=\kappa _{\alpha }\mathbf{1}-\sum_{\mu =1}^{D}\epsilon
^{\mu }N_{\mu \alpha }+O(\epsilon ^{2}).
\end{equation}

The TPCP condition (Eq.~(\ref{cp})) in the lowest order of the parameter $\epsilon$ is given by 
\begin{equation}
\sum_{a}M_{\mu a}^{\dagger }M_{\mu a}=\sum_{\alpha }\left( \kappa
_{\alpha }N_{\mu \alpha }^{\dagger }+\kappa _{\alpha }^{\ast }N_{\mu \alpha
}\right) .
\end{equation}
Due to this relation, the operator $\sum_{\alpha} \kappa _{\alpha }^{\ast }N_{\mu \alpha
} $ can be broken down into a sum of two terms such that
\begin{equation}
\sum_{\alpha }\kappa _{\alpha }^{\ast }N_{\mu \alpha }=\frac{1}{2}\sum_{a}M_{\mu a}^{\dag }M_{\mu a}+iH_{\mu },  \label{3}
\end{equation}
where $H_{\mu }$ are Hermitian operators. Using Eqs.~(\ref{lnc}) and (\ref{3}), the derivative of $\rho (\epsilon )$ at $\epsilon =0$ is evaluated as
\begin{align}
\partial _{\mu }\rho (0)& =\sum_{a=1}^{K_{\mu }}\left[ M_{\mu a}\rho
(0)M_{\mu a}^{\dag }-\frac{1}{2}M_{\mu a}^{\dag }M_{\mu a}\rho (0)-\frac{1}{2}\rho (0)M_{\mu a}^{\dag }M_{\mu a}\right]  \notag \\
& -i[H_{\mu },\ \rho (0)].  \label{lndr}
\end{align}%
The first term of the right hand side indicates a noise effect with
decoherence induced by the environment. On the other hand, the second term expresses a unitary time evolution of the system $S$. Therefore, if the first term vanishes, the purity of the output states does not change in $O(\epsilon )$. In the later discussions, we assume that the operators $M_{\mu a}$ are non-vanishing and thus, in general, the second term does not disappear. This assumption implies that decoherence by the environment inevitably takes place in the system $S$. In this case, the low-noise channel is called dissipative.

\section{ Channel Estimation \label{channel-estimation}}
\ 

We now discuss the attainability of the Cram\'{e}r-Rao inequality for the multi-parameter estimation in the low-noise channel.
Let us consider a pure input state $|\phi \rangle \langle \phi |$ for the 
low-noise channel $\Gamma _{\epsilon }$. 
The output state $\rho (\epsilon )=\Gamma _{\epsilon }[|\phi \rangle \langle \phi |]$ can be diagonalized as 
\begin{equation}
\rho (\epsilon )=\sum_{n=0}^{N-1}p_{n}(\epsilon )|n(\epsilon )\rangle
\langle n(\epsilon )|  \label{sd}
\end{equation}
with an orthonormal basis $\{ | n(\epsilon )\rangle \}$.  Because of a property of the low-noise channel (See Eq.~(\ref{id})), we are able to impose the following boundary conditions, $\rho (0)=|\phi \rangle \langle \phi |$, 
$p_{0}(0)=1,$ and $ p_{n\neq 0}(0)=0$, which imply $|0(0)\rangle =|\phi \rangle .$

The deviation operator $\delta \rho (\epsilon )=\rho (\epsilon )-|\phi
\rangle \langle \phi |$ between the input state and the output state is
evaluated by use of Eq.~(\ref{lndr}) as
\begin{align}
\delta \rho (\epsilon )& =\sum_{\mu }\varepsilon ^{\mu }\partial _{\mu }\rho
(0)+O(\epsilon ^{2})  \notag \\
& =\sum_{\mu }\varepsilon ^{\mu }\sum_{a}\left[ M_{\mu a}|\phi
\rangle \langle \phi |M_{\mu a}^{\dag }-\frac{1}{2}M_{\mu a}^{\dag }M_{\mu
a}|\phi \rangle \langle \phi |-\frac{1}{2}|\phi \rangle \langle \phi |M_{\mu
a}^{\dag }M_{\mu a}\right]  \notag \\
& -i\sum_{\mu }\varepsilon ^{\mu }[H_{\mu },\ |\phi \rangle \langle \phi
|]+O(\epsilon ^{2}).  \label{8}
\end{align}
Here, let us define an $N-1$ dimensional matrix by
\begin{equation}
\Delta (\epsilon )=[\langle n(0)|\delta \rho (\epsilon )|n^{\prime
}(0)\rangle ].  \label{deltam}
\end{equation}
Note that all of the orthonormal vectors $|n(0)\rangle $  $(n=1,\cdots ,N-1)$ are orthogonal to $|\phi \rangle $ by definition. Then we see that the matrix $\Delta (\epsilon )$ is rewritten using Eq.~(\ref{8}) such that
\begin{equation}
\Delta (\epsilon )=\left[ \sum_{\mu a }\epsilon ^{\mu }\langle
n(0)|M_{\mu a }|\phi \rangle \langle \phi |M_{\mu a }^{\dagger
}|n^{\prime }(0)\rangle \right] +O(\epsilon ^{2}).
\end{equation}
The leading term of $\Delta(\epsilon )$ is $O(\epsilon )$, because, as mentioned before, the Kraus operators $M_{\mu a }$ do not vanish due to the assumption of the dissipative low-noise channel. Thus the matrix $\Delta (\epsilon )$
possesses $O(\epsilon )$ eigenvalues. Some eigenvalues of $\Delta (\epsilon
) $ may be higher terms as $O(\epsilon ^{2})$ or exactly zero. However,
the presence of such eigenvalues does not affect the following analysis at all.

The matrix $\Delta(\epsilon)$ becomes a  diagonal matrix with the basis $\{ |n(0)\rangle \}$ in $O(\epsilon)$ and the $N-1$ eigenvalues are  $\delta p_{n}$ defined by $\delta p_{n}(\epsilon)= p_{n}(\epsilon) - p_{n}(0)$, that is,  
\begin{equation}
\delta \rho (\epsilon )|n(0)\rangle =\delta p_{n}(\epsilon )|n(0)\rangle
+O(\epsilon ^{2}),~(n=1,\cdots ,N-1).  \label{eigen}
\end{equation}
It is worthwhile noting that the eigenvalues $\delta p_{n}(\epsilon )$ are
unchanged even if we use a different orthonormal basis $|\widetilde{m}(0)\rangle :=\sum_{n}U_{mn}$ $|n(0)\rangle $ in the definition of $\Delta
(\epsilon )$, where $U_{mn}$ is a unitary matrix. Thus, when one calculate $\delta p_{n}(\epsilon )$, the orthonormal basis $\left\{ |n(0)\rangle
\right\} $ can be arbitrarily fixed in Eq.~(\ref{eigen}). In addition, for $K\leq N-1$,  non-vanishing $\delta p_{n}(\epsilon )$s can be obtained by solving another eigenvalue equation (See Appendix). 
Using the eigenvalues $\delta p_{n}(\epsilon )$, we obtain another expression
of $\rho (\epsilon )$ such that
\begin{equation}
\rho (\epsilon )=|\phi \rangle \langle \phi |+\sum_{\mu }\epsilon ^{\mu
}\sum_{n=0}^{N-1}\left( \partial _{\mu }\delta p_{n}(\epsilon )|n(\epsilon
)\rangle \langle n(\epsilon )|+|\partial _{\mu }0(\epsilon )\rangle \langle
\phi |+|\phi \rangle \langle \partial _{\mu }0(\epsilon )|\right)
+O(\epsilon ^{2}),  \label{r2}
\end{equation}
where $|\partial _{\mu }0(\epsilon )\rangle $ means $\partial _{\mu
}|0(\epsilon )\rangle $. From Eq.~(\ref{r2}), the following relation is
straightforwardly obtained, 
\begin{equation}
\langle n(\epsilon )|\partial _{\mu }\rho (\epsilon )|m(\epsilon
)\rangle =\partial _{\mu }\delta p_{n}(\epsilon )\delta _{nm}+O(\epsilon ),
\label{dr}
\end{equation}
for $n, m \neq0$.

Now we describe the Fisher information matrix for the low noise channel. From Eq.~(\ref{sd}), the SLD's are calculated \cite{hels} as
\begin{equation}
\langle n(\epsilon )|L_{\mu }(\epsilon )|m(\epsilon )\rangle =\frac{2}{p_{n}(\epsilon )+p_{m}(\epsilon )}\langle n(\epsilon )|\partial _{\mu }\rho
(\epsilon )|m(\epsilon )\rangle .  \label{sld}
\end{equation}
Using this, we have the Fisher information matrix $J(\epsilon)=[J_{\mu \nu }(\epsilon )]$ as 
\begin{equation}
J_{\mu \nu }(\epsilon )=\sum_{nm}\langle n(\epsilon )|\partial _{\mu }\rho
(\epsilon )|m(\epsilon )\rangle \frac{2}{p_{n}(\epsilon )+p_{m}(\epsilon )}%
\langle m(\epsilon )|\partial _{\nu }\rho (\epsilon )|n(\epsilon )\rangle .
\label{fim2}
\end{equation}
If all of the eigenvalues $\delta p_{n}(\epsilon )$ are $O(\epsilon )$, the
Fisher information matrix $J(\epsilon )$ is evaluated from Eq.~(\ref{dr}) as 
\begin{equation}
J_{\mu \nu }(\epsilon )=\sum_{n=1}^{N-1}\frac{\partial _{\mu }\delta
p_{n}(\epsilon )\partial _{\nu }\delta p_{n}(\epsilon )}{\delta
p_{n}(\epsilon )}+O(\epsilon ^{0}),  \label{fim}
\end{equation}%
If some of the eigenvalues $\delta p_{n}(\epsilon )$ are $O(\epsilon ^{2})$
or exactly zeros, the terms associated with the $O(\epsilon ^{2})$
eigenvalues in the sum of Eq.~(\ref{fim}) can be neglected. Taking account of
a fact that the leading order of $J(\epsilon )$ is $O(\epsilon ^{-1})$, \ let us define a divergent part of $J(\epsilon )$ by%
\begin{equation}
J_{\mu \nu }^{div}(\epsilon ):=\sum_{n=1}^{N-1}\frac{\partial _{\mu }\delta
p_{n}(\epsilon )\partial _{\nu }\delta p_{n}(\epsilon )}{\delta
p_{n}(\epsilon )}.  \label{jdiv}
\end{equation}%
It is worthwhile noting that the classical Fisher information $J_{\mu \nu
}^{c}(\epsilon )$ defined by
\begin{equation}
J_{\mu \nu }^{c}(\epsilon ):=\sum_{n=0}^{N-1}\frac{\partial _{\mu
}p_{n}(\epsilon )\partial _{\nu }p_{n}(\epsilon )}{p_{n}(\epsilon )}%
=4\sum_{n=0}^{N-1}\partial _{\mu }\sqrt{p_{n}(\epsilon )}\partial _{\nu }%
\sqrt{p_{n}(\epsilon )}
\end{equation}%
coincides with $J_{\mu \nu }^{div}(\epsilon )$ in $O(\epsilon
^{-1})$:
\begin{equation}
J_{\mu \nu }^{div}(\epsilon )=J_{\mu \nu }^{c}(\epsilon )+O(\epsilon ^{0}).
\label{divc}
\end{equation}%
The classical information matrix $J_{\mu \nu }^{c}(\epsilon )$ is an induced
metric of a $D$ dimensional manifold $\emph{M}$ embedded in an $N-1$
dimensional space $\mathit{P~}$defined by $\ \mathit{P}:=\left\{ \left( 
\sqrt{p_{0}},\cdots ,\sqrt{p_{N-1}}\right) |p_{i}\geqq
0,\sum_{i=0}^{N-1}p_{i}=1\right\} $. When $(\epsilon ^{1},\cdots ,\epsilon
^{D})$ locally has a one-to-one correspondence with a point on the manifold $%
\emph{M}$, the matrix $J_{\mu \nu }^{c}(\epsilon )$ has its inverse matrix $%
J^{c,\mu \nu }(\epsilon )$ which is $O(\epsilon )$. If $D>N-1$, some
parameters of $\epsilon $ are redundant and the parameterization of $%
p_{n}(\epsilon )$ by $\epsilon ^{\mu }$ becomes degenerate, that is,
\begin{equation*}
\det_{\mu \nu }\left[ \sum_{n=0}^{N-1}\partial _{\mu }\sqrt{p_{n}(\epsilon )}%
\partial _{\nu }\sqrt{p_{n}(\epsilon )}\right] =0.
\end{equation*}%
Hence, the inverse matrix $J^{c,\mu \nu }(\epsilon )$ does not exist. On the
other hand, when $D\leq N-1$, the inverse matrix $J^{c,\mu \nu }(\epsilon )$
exists for $p_{n}(\epsilon )$ non-degenerate in the neighborhood that $\epsilon=0$. Then, because of this fact and Eq.~(\ref{divc}), the inverse matrix $J^{div.\mu \nu
}(\epsilon )$ of $J_{\mu \nu }^{div}(\epsilon )$ also exists and its order
is $O(\epsilon )$ just as $J^{c,\mu \nu }(\epsilon )$.\ In later analysis,
we assume non-degeneracy of the parameterization;
\begin{equation}
\det_{\mu \nu }\left[ \sum_{n=0}^{N-1}\partial _{\mu }\sqrt{p_{n}(\epsilon )}%
\partial _{\nu }\sqrt{p_{n}(\epsilon )}\right] \neq 0  \label{nondegenerate}
\end{equation}%
for $D\leq N-1$. In this case, the quantum Fisher information $J(\epsilon )$ also has its inverse
matrix $J^{-1}(\epsilon )=[J^{\mu \nu }(\epsilon )]$, whose order is $%
O(\epsilon )$, and satisfies that%
\begin{equation}
J^{\mu \nu }(\epsilon )=J^{div.\mu \nu }(\epsilon )+O(\epsilon ^{2}).
\label{7}
\end{equation}%

Now we shall prove that a locally unbiased estimator, which attains the Cram\'{e}%
r-Rao bound in  $O(\epsilon),$\ can be explicitly constructed.
Firstly, let us introduce commuting Hermitian operators $A_{\mu }(\epsilon )$
given by
\begin{equation}
A_{\mu }(\epsilon ):=\sum_{n=1}^{N-1}\frac{\partial _{\mu }\delta
p_{n}(\epsilon )}{\delta p_{n}(\epsilon )}|n(\epsilon )\rangle \langle
n(\epsilon )|.  \label{ue}
\end{equation}%
If some of $\delta p_{n}(\epsilon )$ are $O(\epsilon ^{2})$, the
corresponding terms are neglected in the sum of Eq.~(\ref{ue}). Their
contravariant operators $A^{\mu}(\epsilon )$ are also defined by%
\begin{equation}
A^{\mu }(\epsilon ):=\sum_{\nu }J^{div.\mu \nu }(\epsilon )A_{\nu }(\epsilon
).  \label{ube}
\end{equation}%
Note that the order of $A^{\mu }(\epsilon )$ is $O(\epsilon ^{0})$. This operator can be written by the spectral decomposition as 
\begin{equation}
A^{\mu }(\epsilon )=\sum_{n}x_{n}^{\mu }(\epsilon )P_{n}(\epsilon ),
\end{equation}%
where $P_{n}(\epsilon )$ are projection operators satisfying $%
\sum_{n}P_{n}(\epsilon )=\mathbf{1}$. Here we define a POVM operator by
\begin{equation}
\Pi _{x}=\sum_{n=1}^{N-1}\prod_{\mu =1}^{D}\delta \left( x^{\mu
}-x_{n}^{\mu }(\epsilon )\right) P_{n}(\epsilon ).
\end{equation}
We adopt $\Pi _{x}$ as an estimator for a measurement to estimate $\epsilon $.
Then it is proven from Eq.~(\ref{dr}) that the estimator satisfies
\begin{eqnarray}
\int x^{\mu }\Tr[\Pi _{x}\rho (\epsilon )]d^{D}x &=&\Tr[A^{\mu }(0)\rho
(\epsilon )] \nonumber \\
&=&\epsilon ^{\mu }+O(\epsilon ^{2}).
\end{eqnarray}%
Thus the estimator $\Pi _{x}$ is locally unbiased for the $\epsilon $
estimation in the vicinity of $\epsilon =0$.  It is also shown that the estimator locally attains the Cram\'{e}r-Rao bound as follows. Using Eq.~(\ref{ue}) and Eq.~(\ref{ube}), we have 
\begin{equation}
\frac{1}{2}\Tr[\rho (\epsilon )\{A^{\mu }(\epsilon ),\ A^{\nu }(\epsilon
)\}]=J^{div.\mu \nu }(\epsilon )+O(\epsilon ^{2}).  \label{5}
\end{equation}
On the other hand, the mean-square-error matrix $V^{\mu \nu }$ for $\Pi _{x}$ is evaluated as
\begin{eqnarray*}
V^{\mu \nu } &=&\int (x^{\mu }-\epsilon ^{\mu })(x^{\nu }-\epsilon ^{\nu
})\Tr[\Pi _{x}\rho (\epsilon )]d^{D}x \\
&=&\frac{1}{2}\Tr[\rho (\epsilon )\{A^{\mu }(\epsilon ),\ A^{\nu }(\epsilon
)\}]+O(\epsilon ^{2}).
\end{eqnarray*}%
Consequently, we see that the Cram\'{e}r-Rao bound is actually
satisfied in  $O(\epsilon )$:
\begin{equation}
V^{\mu \nu }=J^{\mu \nu }(\epsilon )+O(\epsilon ^{2}).
\end{equation}%
This is a main result of this paper. This implies that the lower bound can
always be achieved in the estimation of the dissipative low-noise channel for $D\leq N-1$ if it satisfies Eq.~(\ref{nondegenerate}). Because, generally,
multi-parameter estimation cannot attain the bound, this result is very
significant.

In the case with $D>N-1$, we are able to give another important discussion.
The inverse matrix $J^{-1}(\epsilon )$ of the quantum Fisher information matrix
may exist in this case, even though the classical Fisher information matrix $%
J_{\mu \nu }^{c}(\epsilon)$ does not have its inverse. However, even if the
inverse matrix exists, it is not sufficiently suppressed in general. Its
components are generally $O(\epsilon ^{0})$, not $O(\epsilon )$. Moreover,
the Cram\'{e}r-Rao bound cannot be always attained even in  $O(\epsilon ^{0})$. The disadvantage can be remedied by using the
ancilla-extension of the low-noise channel \cite{hko2}. In fact, the bound
becomes attainable again when $D\leq N^{2}-1$. This is the second result of
this paper.

To see this, let us consider an ancilla system $A$ which has a Hilbert space with the same dimension $N$ as that of the original system $S$. Suppose that the dissipative low-noise channel is extended to $\Gamma _{\epsilon }\otimes id_{A}$, where $id_{A}$ stands for the identity channel on the ancilla system. Entangled states of the composite system $S+A$ \ are available as input states for the channels and a collective measurements is performed for the output states in order to estimate the $D$ parameters $\epsilon ^{\mu }$. In this case, the dimension $N_{S+A}$ of the Hilbert space of $S+A$ is larger than that of $S$ because $N_{S+A}=N^{2}$. Therefore, when the number of the parameters $\epsilon ^{\mu }$ is
less than $N_{S+A}$ ($D\leq N_{S+A}-1$), the above analysis shows that the Cram\'{e}r-Rao bound can be attained in $O(\epsilon)$ assuming that the ancilla-extended channels satisfy Eq.~(\ref{nondegenerate}) for their output states. Consequently, when $N \leq D\leq N^{2}-1$, the ancilla-assisted enhancement effect is able to make the Cram\'{e}r-Rao bound achievable in $O(\epsilon )$.

\section{Examples}
\

Let us consider a two-parameter dissipative low-noise channel $\Gamma _{\epsilon }$ for a system $S$ with $N = 3$ and $D=2$ which is a simple example satisfying $K \leq N-1$, and focus on the case with $K_{1}=K_{2}=1$. Suppose that an input state $|\phi \rangle$ go through the channel. We apply the method described in Appendix to solve the eigenvalues of the $\delta p_{\pm}(\epsilon)$ of the output state. In this present system, the matrix $\Lambda (\epsilon)$ defined by Eq.~(\ref{lambda}) can be written as 
\begin{equation*}
 \Lambda (\epsilon )=\left[ \sqrt{\epsilon ^{\mu }}\langle \phi |\left(
M_{\mu }-\langle \phi |M_{\mu }|\phi \rangle \right) \left( M_{\nu }-\langle
\phi |M_{\nu }|\phi \rangle \right) |\phi \rangle \sqrt{\epsilon ^{\nu }}%
\right],
\end{equation*}
where $M_{\mu}$ denotes $M_{\mu 1}$ with $\mu =1,2,3.$ The following eigenvalues is obtained by solving Eq.~(\ref{det}),
\begin{eqnarray*}
&&\lefteqn{\delta p_{\pm }(\epsilon)} \\
&&=\frac{1}{2}\left[ \epsilon ^{1}\delta M_{11}+\epsilon ^{2}\delta
M_{22}\pm \sqrt{\left( \epsilon ^{1}\delta M_{11}-\epsilon ^{2}\delta
M_{22}\right) ^{2}+4\epsilon ^{1}\epsilon ^{2}\delta M_{12}\delta M_{21}}%
\right] \\
&&=O(\epsilon ),
\end{eqnarray*}%
where $\delta M_{\mu \nu }$ are variance-matrix elements defined by 
\begin{equation*}
\left[ 
\begin{array}{cc}
\delta M_{11} & \delta M_{12} \\ 
\delta M_{21} & \delta M_{22}%
\end{array}
\right] =\left[ \langle \phi |\left( M_{\mu }-\langle \phi |M_{\mu }|\phi
\rangle \right) \left( M_{\nu }-\langle \phi |M_{\nu }|\phi \rangle \right)
|\phi \rangle \right] .
\end{equation*}
The inverse of the Fisher information matrix can be explicitly calculated as
\begin{eqnarray*}
J^{11}&=&\frac{\left( \epsilon ^{1}\right) ^{3}\delta M_{11}\det \delta
M+\left( \epsilon ^{1}\right) ^{2}\epsilon ^{2}\delta M_{22}\left[ 3\delta
M_{12}\delta M_{21}-2\delta M_{11}\delta M_{22}\right] +\epsilon ^{1}\left(
\epsilon ^{2}\right) ^{2}\delta M_{22}^{3}}{\left( \delta M_{11}\delta
M_{22}-\delta M_{12}\delta M_{21}\right) \left( \epsilon ^{1}\delta
M_{11}-\epsilon ^{2}\delta M_{22}\right) ^{2}}\\
J^{22}&=&\frac{\left( \epsilon ^{2}\right) ^{3}\delta M_{22}\det \delta
M+\left( \epsilon ^{2}\right) ^{2}\epsilon ^{1}\delta M_{11}\left[ 3\delta
M_{12}\delta M_{21}-2\delta M_{11}\delta M_{22}\right] +\epsilon ^{2}\left(
\epsilon ^{1}\right) ^{2}\delta M_{11}^{3}}{\left( \delta M_{11}\delta
M_{22}-\delta M_{12}\delta M_{21}\right) \left( \epsilon ^{1}\delta
M_{11}-\epsilon ^{2}\delta M_{22}\right) ^{2}}
\end{eqnarray*}
\begin{equation}
J^{12}=J^{21}=-\epsilon ^{1}\epsilon ^{2}\frac{\delta M_{12}\delta M_{21}}{%
\delta M_{11}\delta M_{22}-\delta M_{12}\delta M_{21}}\frac{\epsilon
^{1}\delta M_{11}+\epsilon ^{2}\delta M_{22}}{\left( \epsilon ^{1}\delta
M_{11}-\epsilon ^{2}\delta M_{22}\right) ^{2}}.
\end{equation}%
Hence, for input states which satisfy
\begin{eqnarray}
\delta M_{11}\delta M_{22}-\delta M_{12}\delta M_{21} &=&O(\epsilon^{0}), \nonumber\\
n^{1}\delta M_{11}-n^{2}\delta M_{22} &=&O(\epsilon^{0}), \nonumber
\end{eqnarray}
with $n^{\mu } := \epsilon^{\mu}/\epsilon$, the inverse matrix $J^{-1}$ behaves as $O(\epsilon)$.  Thus, the Cram\'{e}r-Rao lower bound can be attained in $O(\epsilon)$, since one can make an locally unbiased estimator to attain the bound by following the procedure in the previous section.

Next, let us give another example for $N=2$ and $D=2$ to show the ancilla-assisted improvement as mentioned in Sec.~\ref{channel-estimation}.  Now suppose that an input state $\rho _{in}$
goes through a dissipative low-noise channel $\tilde{\Gamma}_{\epsilon}$
defined by
\begin{equation*}
\tilde{\Gamma} _{\epsilon}\left[ \rho _{in}\right] =\left( 1-\epsilon
^{1}-\epsilon ^{2}\right) \rho _{in}+\epsilon ^{1}\sigma _{x}\rho
_{in}\sigma _{x}+\epsilon ^{2}\sigma _{z}\rho _{in}\sigma _{z},
\end{equation*}%
where $\epsilon ^{\mu }$ are unknown low-noise parameters and $\sigma _{x}$ and $\sigma _{y}$ are the Pauli matrices. Its output state $\rho (\epsilon )$
can be expressed by a Bloch representation as
\begin{equation}
\tilde{\rho} (\epsilon)=\tilde{\Gamma}_{\epsilon}\left[ \rho _{in}\right] =\frac{1%
}{2}+\frac{1}{2}\vec{y}\cdot \vec{\sigma},
\end{equation}%
where $\vec{y}$ is a Bloch vector and $\vec{\sigma}=\left( \sigma
_{x},\sigma _{y},\sigma _{z}\right)$. The SLD for the output states is exactly solved and written as
\begin{equation}
L_{\mu }=l_{o,\mu }+\vec{l}_{\mu }\cdot \vec{\sigma},
\end{equation}
where
\begin{equation}
l_{o,\mu }=-\frac{1}{2}\frac{\partial _{\mu }|\vec{y}|^{2}}{1-|\vec{y}|^{2}}\nonumber,
\end{equation}
and
\begin{equation}
\vec{l}_{\mu }=\partial _{\mu }\vec{y}+\frac{1}{2}\frac{\partial _{\mu }|%
\vec{y}|^{2}}{1-|\vec{y}|^{2}}\vec{y}\nonumber.
\end{equation}%
The Fisher information matrix $J_{\mu \nu }$ is directly calculated as
\begin{equation}
J_{\mu \nu }=\partial _{\mu }\vec{y}\cdot \partial _{\nu }\vec{y}+\frac{1}{4}%
\frac{\partial _{\mu }|\vec{y}|^{2}\partial _{\nu }|\vec{y}|^{2}}{1-|\vec{y}%
|^{2}}.
\end{equation}%
Defining $\delta p=4(1-|\vec{y}|^{2})$ which vanishes when $\epsilon =0$,
the inverse matrix $J^{\mu \nu }$ is calculated as 
\begin{eqnarray}
J^{11}&=&\frac{|\partial _{2}\vec{y}|^{2}+\frac{1}{\delta p}\left( \partial
_{2}|\vec{y}|^{2}\right) ^{2}}{|\partial _{1}\vec{y}|^{2}|\partial _{2}\vec{y}|^{2}-(\partial _{1}\vec{y}\cdot \partial _{2}\vec{y})^{2}+\frac{1}{\delta p}\left| \partial _{2}|\vec{y}|^{2}\partial _{1}\vec{y}-\partial _{1}|\vec{y}
|^{2}\partial _{2}\vec{y}\right|^{2}},\nonumber \\
J^{22}&=&\frac{|\partial _{1}\vec{y}|^{2}+\frac{1}{\delta p}\left( \partial
_{1}|\vec{y}|^{2}\right) ^{2}}{|\partial _{1}\vec{y}|^{2}|\partial _{2}\vec{y%
}|^{2}-(\partial _{1}\vec{y}\cdot \partial _{2}\vec{y})^{2}+\frac{1}{\delta p%
}\left| \partial _{2}|\vec{y}|^{2}\partial _{1}\vec{y}-\partial _{1}|\vec{y}%
|^{2}\partial _{2}\vec{y}\right|^{2}}, \nonumber \\
J^{12}&=&-\frac{\partial _{1}\vec{y}\cdot \partial _{2}\vec{y}+\frac{1}{\delta
p}\partial _{1}|\vec{y}|^{2}\partial _{2}|\vec{y}|^{2}}{|\partial _{1}\vec{y}%
|^{2}|\partial _{2}\vec{y}|^{2}-(\partial _{1}\vec{y}\cdot \partial _{2}\vec{%
y})^{2}+\frac{1}{\delta p}\left| \partial _{2}|\vec{y}|^{2}\partial _{1}\vec{%
y}-\partial _{1}|\vec{y}|^{2}\partial _{2}\vec{y}\right|^{2}}.
\end{eqnarray}
By solving the eigenvalue equation of $J^{-1}$, we can show that one of the eigenvalues
is $O(\epsilon)$, but the other is $O(\epsilon^{0})$, namely,  
\begin{equation}
diag[J^{-1}]=\left[ 
\begin{array}{cc}
O(\epsilon^{0}) & 0 \\ 
0 & O(\epsilon )%
\end{array}%
\right] .
\end{equation}%
Therefore, due to the $O(\epsilon^{0})$ lower bound, it is impossible to estimate the
parameters with $O(\epsilon ^{1/2})$ precision. This property is briefly illustrated when $\epsilon=0$ at which the inverse matrix can be written as
\begin{equation*}
J^{-1}(0)=\frac{1}{\Phi }\left[ 
\begin{array}{c}
\partial _{2}|\vec{y}|^{2} \\ 
-\partial _{1}|\vec{y}|^{2}%
\end{array}%
\right] \left[ \partial _{2}|\vec{y}|^{2}\ -\partial _{1}|\vec{y}|^{2}\right]
,
\end{equation*}%
with $\Phi :=\left\vert \partial _{2}|\vec{y}|^{2}\partial _{1}\vec{y}%
-\partial _{1}|\vec{y}|^{2}\partial _{2}\vec{y}\right\vert ^{2}$. It is easily seen that $J^{-1}$ has a zero eigenvalue with an eigenvector $\left[ 
\begin{array}{cc}
\partial _{1}|\vec{y}|^{2} & \partial _{2}|\vec{y}|^{2}%
\end{array}%
\right] ^{T}$, that is, 
\begin{equation*}
J^{-1}(0)\left[ 
\begin{array}{c}
\partial _{1}|\vec{y}|^{2} \\ 
\partial _{2}|\vec{y}|^{2}%
\end{array}%
\right] =\vec{0}.
\end{equation*}%
It turns out in $O(\epsilon ^{0})$ that the vector $\left[ 
\begin{array}{cc}
\partial _{1}|\vec{y}|^{2} & \partial _{2}|\vec{y}|^{2}%
\end{array}%
\right] ^{T}$ corresponds to the eigenvector with the eigenvalue of $O(\epsilon)$ and the other eigenvector $\left[ 
\begin{array}{cc}
\partial _{2}|\vec{y}|^{2} & -\partial _{1}|\vec{y}|^{2}%
\end{array}%
\right] ^{T}$ has the eigenvalue of $O(\epsilon^{0})$.

In order to attain the Cram\'{e}r-Rao bound of $O(\epsilon)$, let us consider an ancilla-extension of the channel, $\tilde{\Gamma}_{\epsilon
}\otimes id_{A}$, and its entangled input state $| \Psi \rangle$ described by 
\begin{equation}
|\Psi \rangle =\frac{1}{\sqrt{2}}(|++\rangle +|--\rangle),
\end{equation}
where $\sigma _{z}|\pm \rangle =\pm |\pm \rangle $. Then, the output state is given by $(\Gamma _{\epsilon }\otimes id_{A})[|\Psi\rangle \langle \Psi |]$. In order to evaluate the eigenvalues $\delta p_{n}(\epsilon )$, let us
define three vectors
\begin{eqnarray}
|1\rangle &=&\frac{1}{\sqrt{2}}(|++\rangle -|--\rangle), \nonumber\\
|2\rangle &=&|+-\rangle , \nonumber\\
|3\rangle &=&|-+\rangle .
\end{eqnarray}%
Note that these vectors are mutually orthogonal and all of them are orthogonal to the input state $|\Psi\rangle$. Thus it is possible to represent the matrix $\Delta (\epsilon)$ (See Eq.~(\ref{deltam})) with a basis $\{|1\rangle , |2\rangle , |3\rangle  \}$ as
\begin{equation}
\Delta (\epsilon )=\left[ 
\begin{array}{ccc}
\epsilon ^{2} & 0 & 0 \\ 
0 & \epsilon ^{1}/2 & \epsilon ^{1}/2  \\ 
0 & \epsilon ^{1}/2 & \epsilon ^{1}/2%
\end{array}
\right].
\end{equation}
By estimating eigenvalues of $\Delta (\epsilon )$, it turns out that all
the three eigenvalues behave as $O(\epsilon )$. In fact, the three eigenvalues are evaluated as
\begin{eqnarray}
\delta p_{1}(\epsilon ) &=&\epsilon ^{2}, \nonumber\\
\delta p_{2}(\epsilon ) &=&\epsilon ^{1}, \nonumber \\
\delta p_{3}(\epsilon ) &=&0.
\end{eqnarray}%
Substituting $\delta p_{n}(\epsilon )$ into Eq.~(\ref{fim}) gives the Fisher
information matrix such that 
\begin{equation*}
\left[ J_{\mu \nu }\right] =\left[ 
\begin{array}{cc}
\frac{1}{\epsilon ^{1}} & 0 \\ 
0 & \frac{1}{\epsilon ^{2}}%
\end{array}%
\right] +O(\epsilon ^{0}).
\end{equation*}
Here the term $\partial _{\mu }\delta p_{3}(\epsilon )\partial _{\nu }\delta
p_{3}(\epsilon )/\delta p_{3}(\epsilon )$ in Eq.~(\ref{fim}) has been set to
zero. It can be verified that a mapping onto the manifold $\emph{M}$ defined
by 
\begin{equation*}
\left( \sqrt{p_{0}},\sqrt{p_{1}},\sqrt{p_{2}},\sqrt{p_{3}}\right) :=\left( 
\sqrt{1-\delta p_{1}(\epsilon )-\delta p_{2}(\epsilon )-\delta
p_{3}(\epsilon )},\sqrt{\delta p_{1}(\epsilon )},\sqrt{\delta p_{2}(\epsilon
)},\sqrt{\delta p_{3}(\epsilon )}\right)
\end{equation*}
is non-degenerate;
\begin{equation*}
\det_{{}}\left[ 
\begin{array}{cc}
\sum_{n=0}^{3}\partial _{1}\sqrt{p_{n}(\epsilon )}\partial _{1}\sqrt{%
p_{n}(\epsilon )} & \sum_{n=0}^{3}\partial _{1}\sqrt{p_{n}(\epsilon )}%
\partial _{2}\sqrt{p_{n}(\epsilon )} \\ 
\sum_{n=0}^{3}\partial _{2}\sqrt{p_{n}(\epsilon )}\partial _{1}\sqrt{%
p_{n}(\epsilon )} & \sum_{n=0}^{3}\partial _{2}\sqrt{p_{n}(\epsilon )}%
\partial _{2}\sqrt{p_{n}(\epsilon )}%
\end{array}%
\right] \neq 0.
\end{equation*}%
This non-degeneracy guarantees the existence of $\left[ J^{\mu \nu }\right] $%
. The matrix $\left[ J^{\mu \nu }\right] $ is suppressed as $O(\epsilon )$.
In fact, $\left[ J^{\mu \nu }\right] $ is given by
\begin{equation*}
\left[ J^{\mu \nu }\right] =\left[ 
\begin{array}{cc}
\epsilon ^{1} & 0 \\ 
0 & \epsilon ^{2}%
\end{array}%
\right] +O(\epsilon ^{2}).
\end{equation*}%
Calculation of the projective operators $\left\{ |n(0)\rangle \langle
n(0)|~|n=0,1,2,3\right\} $ is simple and results in

\begin{eqnarray*}
|0(0)\rangle \langle 0(0)| &=&|\Psi \rangle \langle \Psi |, \\
|1(0)\rangle \langle 1(0)| &=&\frac{1}{2}\left( |++\rangle -|--\rangle \right) \left( \langle ++|-\langle --|\right) ,
\\
|2(0)\rangle \langle 2(0)| &=&\frac{1}{2}\left( |+-\rangle +|-+\rangle \right) \left( \langle +-|+\langle -+|\right) ,
\\
|3(0)\rangle \langle 3(0)| &=& \frac{1}{2}\left( |+-\rangle -|-+\rangle \right) \left( \langle +-|- \langle -+|\right) .
\end{eqnarray*}
Adopting the estimator, the ancilla-extension of the channel really makes the
Cram\'{e}r-Rao bound achievable in $O(\epsilon )$.

\bigskip\ \newline
\textbf{Acknowledgement}\newline

\bigskip

We would like to thank Masanao Ozawa for useful discussions in the first
stage of this research. This research was partially supported by the SCOPE
project of the MIC.

\appendix
\section*{Appendix}
Here, we shall show that the non-vanishing $\delta p_{n}(\epsilon )$ satisfy
\begin{equation}
\det \left[ \delta p_{n}(\epsilon )\mathbf{1}_{K}-\Lambda (\epsilon )\right]
=0,  \label{det}
\end{equation}
for $K \leq N-1$, where the matrix $\Lambda$ is a $K\times K$ Hermite matrix defined by 
\begin{eqnarray}
\Lambda (\epsilon )&=&\left[ \Lambda ^{\mu a, \nu b}(\epsilon )\right] \nonumber\\ &=&\left[ 
\sqrt{\epsilon ^{\mu }}\left( \langle \phi |\left( M_{\mu a}^{\dagger
}-\langle \phi |M_{\mu a}^{\dagger }|\phi \rangle \right) \left( M_{\nu
b}-\langle \phi |M_{\nu b}|\phi \rangle \right) |\phi \rangle \right) \sqrt{%
\epsilon ^{\nu }}\right].  \label{lambda}
\end{eqnarray}
This means that one can evaluate $\delta p_{n}(\epsilon )$ by solving Eq.~(\ref{det}).


The proof of Eq.~(\ref{det}) is as follows. We begin with a function $f(p)$ defined by 
\begin{equation}
f(p):=\det \left[ p\mathbf{1}_{N-1}-\Delta (\epsilon )\right].
\end{equation}
This function can be expressed by products of $\Tr\left[ \Delta (\epsilon )^{k}\right] $
as
\begin{eqnarray}
f(p) &=&p^{N-1}-p^{N-2}\Tr\Delta (\epsilon )+\cdots \nonumber\\
&=&p^{N-1}+\sum_{n=1}^{N-1}p^{N-1-n}C_{n}\left( \Tr\left[ \Delta
(\epsilon )^{n}\right] ,\Tr\left[ \Delta (\epsilon )^{n-1}\right] ,\cdots ,\Tr%
\left[ \Delta (\epsilon )\right] \right).\nonumber \\ \label{fp}
\end{eqnarray}
where $\{C_{n}(\cdot)\}$ denote some coefficients expressed as a function of $\Tr[ \Delta(\epsilon)^{k}]$ with $k=1,2,\cdots, n$. In Eq.~(\ref{fp}), the trace of $\Delta (\epsilon)^{k}$ is calculated by use of the relation
$
\sum_{n=1}^{N-1}|n(0)\rangle \langle n(0)|=\mathbf{1}_{N}-|\phi
\rangle \langle \phi |$ as 
\begin{eqnarray*}
&&\lefteqn{\Tr\left[ \Delta (\epsilon )^{k}\right]} \\
&&=\sum_{\mu _{1}\cdots \mu _{k}}\sum_{a^{1}\cdots
a^{k}}\epsilon ^{\mu _{1}}\cdots \epsilon ^{\mu _{k}}  
\Tr[ M_{\mu^{1}a^{1}}|\phi \rangle \langle \phi |M_{\mu ^{1}a^{1}}^{\dag }\left( 
\mathbf{1}_{N}-|\phi \rangle \langle \phi |\right) \\ 
&& \hspace{50mm} \cdots M_{\mu
^{k}a^{k}}|\phi \rangle \langle \phi |M_{\mu ^{k}a^{k}}^{\dag }\left( 
\mathbf{1}_{N}-|\phi \rangle \langle \phi |\right) ] \\
&&=\sum_{\mu _{1}\cdots \mu _{k}}\sum_{a^{1}\cdots a^{k}} \Big[ \sqrt{\epsilon ^{\mu _{k}}}\langle \phi |M_{\mu ^{k}a^{k}}^{\dag }\left( \mathbf{1}_{N}-|\phi \rangle \langle \phi |\right) M_{\mu ^{1}a^{1}}|\phi \rangle \sqrt{\epsilon ^{\mu _{1}}} \\ && \hspace{35mm} \cdots \sqrt{\epsilon ^{\mu _{k-1}}}\langle \phi |M_{\mu
^{k-1}a^{k-1}}^{\dag }\left( \mathbf{1}_{N}-|\phi \rangle \langle \phi
|\right) M_{\mu ^{k}a^{k}}|\phi \rangle \sqrt{\epsilon ^{\mu _{k}}}%
\Big] \\
&&=\Tr\left[ \Lambda (\epsilon )^{k}\right] .
\end{eqnarray*}%
Thus, for $K\leq N-1$, we obtain
\begin{eqnarray*}
f(p) &=&p^{N-1}+\sum_{n=1}^{N-1}p^{N-1-n}C_{n}\left( \Tr\left[
\Lambda (\epsilon )^{n}\right] ,\Tr\left[ \Lambda (\epsilon )^{n-1}\right]
,\cdots ,\Tr\left[ \Lambda (\epsilon )\right] \right) \\
&=&p^{N-1}+\sum_{n=1}^{K}p^{N-1-n}C_{n}\left( \Tr\left[ \Lambda
(\epsilon )^{n}\right] ,\Tr\left[ \Lambda (\epsilon )^{n-1}\right] ,\cdots ,\Tr%
\left[ \Lambda (\epsilon )\right] \right) \\
&=&p^{N-1-K}\left[ p^{K}+\sum_{n=1}^{K}p^{K-n}C_{n}\left( \Tr\left[
\Lambda (\epsilon )^{n}\right] ,\Tr\left[ \Lambda (\epsilon )^{n-1}\right]
,\cdots ,\Tr\left[ \Lambda (\epsilon )\right] \right) \right] \\
&=&p^{N-1-K}\det \left[ p\mathbf{1}_{K}-\Lambda (\epsilon )\right] .
\end{eqnarray*}%
Consequently, applying this result to the eigenvalue equation for $\Delta(\epsilon)$ gives
\begin{equation*}
\det \left[ \delta p_{n}(\epsilon )\mathbf{1}_{K}-\Lambda (\epsilon )\right]
=0,
\end{equation*}%
which completes the proof.

\end{document}